\documentclass[12pt,aps,floats]{revtex4}
\usepackage{blindtext}
\usepackage{graphicx}
\usepackage{color}
\usepackage{amsmath}  % needed for \tfrac, \bmatrix, etc.
\usepackage{amssymb}
\usepackage{amsfonts} % needed for bold Greek, Fraktur, and blackboard bold
\usepackage{graphicx} % needed for figures
\usepackage{xcolor}
\usepackage{color}
\usepackage{caption}
\usepackage[toc,page]{appendix} 
\usepackage{amsmath}
\usepackage{amssymb}

\newcommand{\simlt}
{\lower.5ex\hbox{\ltsima}}
\newcommand{\simgt}
{\lower.5ex\hbox{\gtsima}}

\begin{document}

\title{The Franson experiment as an example of spontaneous breaking of time-translation symmetry}

\author{David H. Oaknin}
\email{d1306av@gmail.com}
\affiliation{%
Rafael Ltd., IL-31021 Haifa, Israel
}%

\begin{abstract}
We describe an explicit statistical model of local hidden variables that reproduces the predictions of quantum mechanics for the ideal Franson experiment and sheds light on the physical mechanisms that might be involved in the actual experiment. The crux of our model is the spontaneous breaking of the time-translation gauge symmetry by the hidden configurations of the pairs of photons locked in time and energy involved in the experiment, which acquire a non-zero geometric phase through certain cyclic transformations. 
\end{abstract}

\date{\today}

\maketitle

\section{Introduction}

It is a widely accepted wisdom that quantum phenomena cannot be fully described within the framework of any physical theory that shares the same notions of reality and relativistic causality that we acknowledge as a given in our classical descriptions of the macroscopic world \cite{Wiseman}. This wisdom is precisely formulated through the Bell theorem on the attainable correlations between the outcomes of polarization measurements performed 
on pairs of photons prepared in a singlet polarization state \cite{EPR,Bohm}. The theorem draws a solid line (the Bell inequality) that allows to experimentally discriminate between the predictions of quantum mechanics for these correlations and those of models of local hidden variables that fulfill certain physically intuitive requirements \cite{Bell,Bell2,CHSH,Fine}. 

In a series of recent papers we have shown, however, that the proof of the Bell theorem relies crucially on a subtle implicit assumption that is not required by fundamental physical principles and, therefore, the Bell inequality does not necessarily hold for models of local hidden variables that do not comply with the said unjustified assumption \cite{david,david0,david1}. In consequence, such models cannot be ruled out by the experimental evidence for the violation of the inequality \cite{Aspect,Weihs,Hansen}.

The Franson experiment is often regarded as an alternative demonstration of the impossibility of describing  quantum phenomena within the framework of any local model of hidden variables \cite{Franson,Franson1,Belinskii,Kwiat1}. As in the case of the Bell experiment, it has been shown that certain features of the predictions of quantum mechanics for the Franson experiment cannot be reproduced within the framework of any model of local hidden variables that shares certain intuitive requirements \cite{Kwiat2,Su,Larsson,Aerts}.

In this paper we argue, however, that as in the case of the Bell experiment the models of local hidden variables whose predictions for the Franson experiment can be distinguished from those of quantum mechanics all share an assumption that is not required by any fundamental principle. Indeed, we explicitly describe a model of local hidden variables that does not comply with the disputed assumption and, hence, it successfully reproduces the predictions of quantum mechanics for the experiment. The crux of our model is the spontaneous breaking of the time-translation gauge symmetry by the hidden configurations of the pairs of photons locked in time and energy involved in the experiment, which acquire due to a holonomy a non-zero geometric phase through certain cyclic coordinate transformations. {\color{black} Let us stress that the gauge symmetry is spontaneously broken when each one of the possible hidden configurations of the pair of photons is considered separately (that is, for every single realization of the experiment), but it is statistically restored over the whole population of all possible hidden configurations (that is, over a long sequence of repetitions). Thus, the expected average correlations do only depend on the gauge-independent physical parameters that describe the settings of the experiment, in agreement with Elitzur's theorem \cite{Elitzur1975}}. The model discussed here for the Franson experiment closely resembles the model of hidden variables introduced in \cite{david,david0,david1} for the Bell experiment.

The paper is organized as follows. In section II we present a (somewhat simplistic) description of the ideal Franson experiment within the framework of quantum mechanics and summarize its predictions {\color{black} (see \cite{Belinskii} for a general review of optical tests of the Bell inequality)}. In section III we present a detailed description of the setup of an actual experiment. In section IV we describe a explicit model of local hidden variables that reproduces the predictions and the collected experimental data described in the two previous sections. Finally, in section V we discuss and summarize our findings.

\section{The Franson experiment}

In a Franson experiment a source produces pairs of photons, $A$ and $B$, whose state is described by a wavefunction of the form
\begin{equation}
\label{wavefunction}
|\Psi\rangle = \frac{|\xi_1\rangle^{(A)} + e^{i \phi_A} |\xi_2\rangle^{(A)}}{\sqrt{2}} \otimes \frac{|\xi_1\rangle^{(B)} + e^{i \phi_B} |\xi_2\rangle^{(B)}}{\sqrt{2}},
\end{equation}
where $\{|\xi_1\rangle, |\xi_2\rangle\}^{(A,B)}$ are orthonormal bases in their respective single-particle Hilbert spaces, 
\begin{equation}
{}^{(A)}\langle \xi_1|\xi_2\rangle^{(A)} = {}^{(B)}\langle \xi_1|\xi_2\rangle^{(B)} = 0,
\end{equation}
and $\phi_A$, $\phi_B$ are phases that, in principle, can be controlled and set at will. 

\begin{figure}
%\begin{center}
\includegraphics[width=12cm]{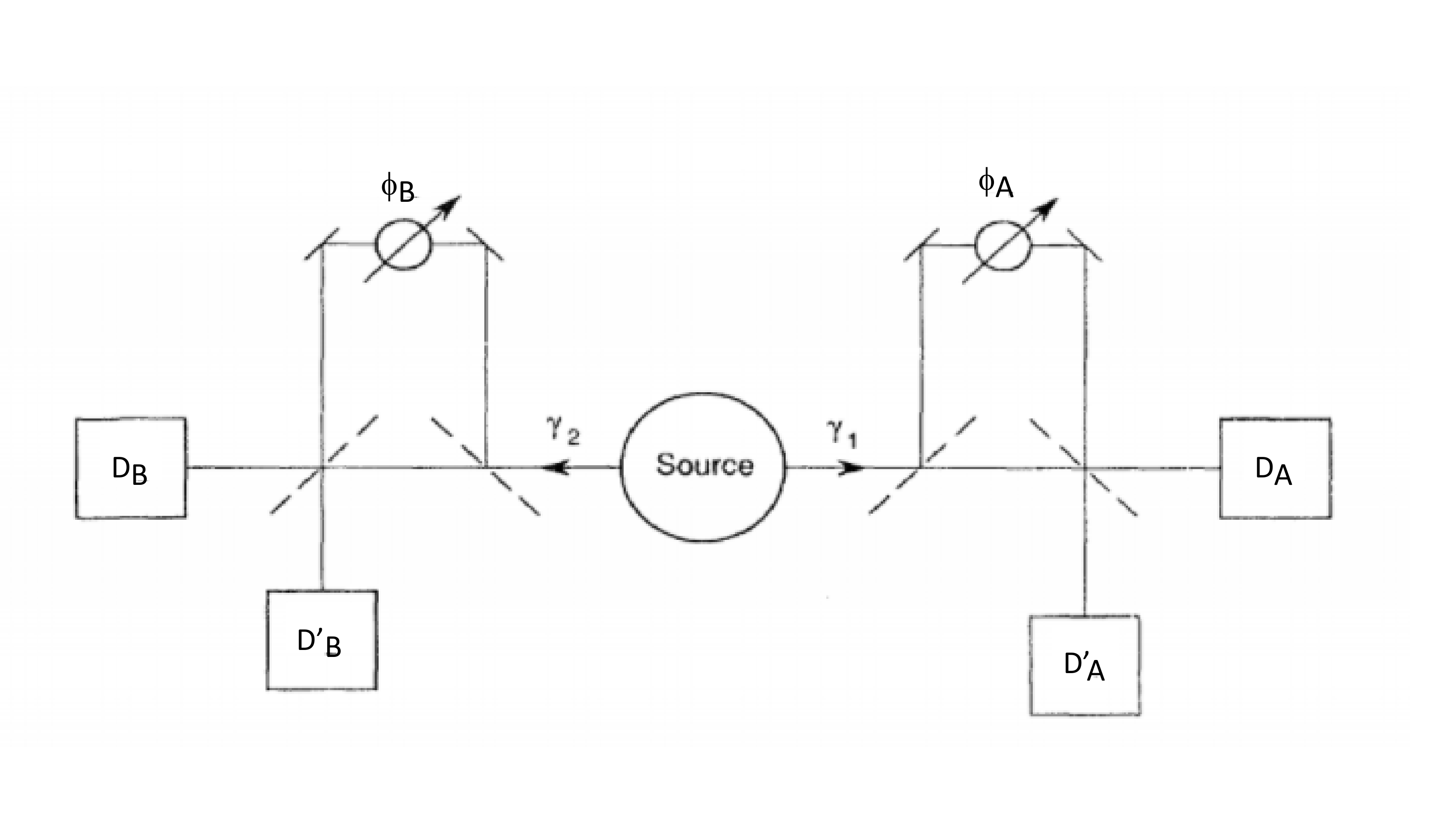}
%\end{center}
\caption{Schematic set-up for an ideal Franson experiment. {\color{black} A source produces pairs of photons locked in time and energy, which are then sent through two unbalanced, perfectly calibrated,  Mach-Zender interferometers. At their exit, each one of the photons is registered by either one of two detectors located at their corresponding ends  ($D_A$, $D_{A'}$ and $D_B$, $D_{B'}$, respectively), which record their times of arrival.} This figure has been taken and reproduced here from reference \cite{Franson}.}
\label{fig:Franson}
\end{figure}

A projective measurement is then performed on the pair of photons along the orthonormal basis in their joint Hilbert space $\left\{|c_1\rangle, |c_2\rangle, |c_3\rangle, |c_4\rangle\right\}$ defined by the vectors

\begin{eqnarray}
\nonumber
|c_1\rangle, & = & \frac{1}{\sqrt{2}}\left(|\xi_1\rangle^{(A)} \otimes |\xi_1\rangle^{(B)} + i |\xi_2\rangle^{(A)} \otimes |\xi_2\rangle^{(B)}\right), \\
\nonumber
|c_2\rangle, & = & \frac{1}{\sqrt{2}}\left(|\xi_1\rangle^{(A)} \otimes |\xi_1\rangle^{(B)} - i |\xi_2\rangle^{(A)} \otimes |\xi_2\rangle^{(B)}\right), \\
\nonumber
|c_3\rangle, & = & |\xi_1\rangle^{(A)} \otimes |\xi_2\rangle^{(B)}, \\
|c_4\rangle, & = & |\xi_2\rangle^{(A)} \otimes |\xi_1\rangle^{(B)},
\end{eqnarray}
so that the probabilities $p_i=\langle\Psi|c_i\rangle\langle c_i|\Psi\rangle$ for each one of the four possible outcomes are given, respectively, by:
\begin{eqnarray}
\nonumber
p_1 & = & \frac{1}{4} \left[1 + \cos(\phi_A+\phi_B-\frac{\pi}{2})\right], \hspace{0.5in} p_3 = \frac{1}{4}, \\
\label{probabilities}
p_2 & = & \frac{1}{4} \left[1 - \cos(\phi_A+\phi_B-\frac{\pi}{2})\right], \hspace{0.5in} p_4 = \frac{1}{4}. 
%p_3 & = & \frac{1}{4}, \\
%p_4 & = & \frac{1}{4},
\end{eqnarray}
Outcomes $\#1$ and $\#2$, to which we shall refer as simultaneous events for reasons that will be clear later on, account for half of all the events, while outcomes $\#3$ and $\#4$, to which we shall refer as non-simultaneous events, account for the other half. 
The experiment is schematically described in Fig. \ref{fig:Franson}. This figure is a reproduction of Fig. 1 of reference \cite{Franson}.

We are interested here in the pattern of interference fringes shown by the probabilities of the simultaneous events, $p_1$ and $p_2$, as a function of the total phase $\Delta=\phi_A+\phi_B$. In particular, the probability $p_1$ is equal (up to a normalization factor) to the probability of 'equal' outcomes - either $(-1,-1)$ or $(+1,+1)$ - at the two polarization measurement devices in a Bell experiment with photons prepared in a singlet polarization state, while the probability $p_2$ is equal (up to the same normalization factor as above) to the probability for 'non-equal' outcomes - either $(+1,-1)$ or $(-1,+1)$ - in the two devices. Hence, following the Bell theorem, it is claimed that these probabilities cannot be reproduced within the framework of any model of local hidden variables \cite{Kwiat2,Su,Larsson}.

Notwithstanding, some authors have raised questions regarding the origin of the claimed 'non-classical' features of the Franson experiment \cite{Ham}, since the pairs of photons are initially prepared in a separable state and they become entangled only when they both are measured, well after they have left their source and do not further interact with each other \cite{Kwiat1}. In this paper we explore these and other questions with the help of an explicit model of local hidden variables that reproduces the predictions of quantum mechanics summarized in (\ref{probabilities}).

Before we proceed we make the following important observation. The orthonormal single-particle eigenstates $|\xi_1\rangle^{(A,B)}$, $|\xi_2\rangle^{(A,B)}$ are defined each up to a global phase (as any normalized eigenvector of any linear operator) and, therefore, the phases $\phi_A$ and $\phi_B$ in the wavefunction (\ref{wavefunction}) have not been properly defined yet. In order to do so it is necessary to set an arbitrary setting of the actual experiment as a reference and measure the resulting probabilities. This reference setting, thus, fixes a reference value for the phase $\phi_A+\phi_B$, with respect to which we can properly define a subsequent change. On the other hand, in the above description the phase difference $\phi_A-\phi_B$ cannot be properly defined and it is, therefore, a spurious degree of freedom, which we can set to $\phi_A-\phi_B=0$. In other words, in the quantum description the setting of the experiment is actually described by a single physical parameter, $\phi_A+\phi_B$, rather than two independent parameters, $\phi_A$ and $\phi_B$.  

\section{The actual experimental set-up}

\begin{figure}
%\begin{center}
\includegraphics[width=12cm]{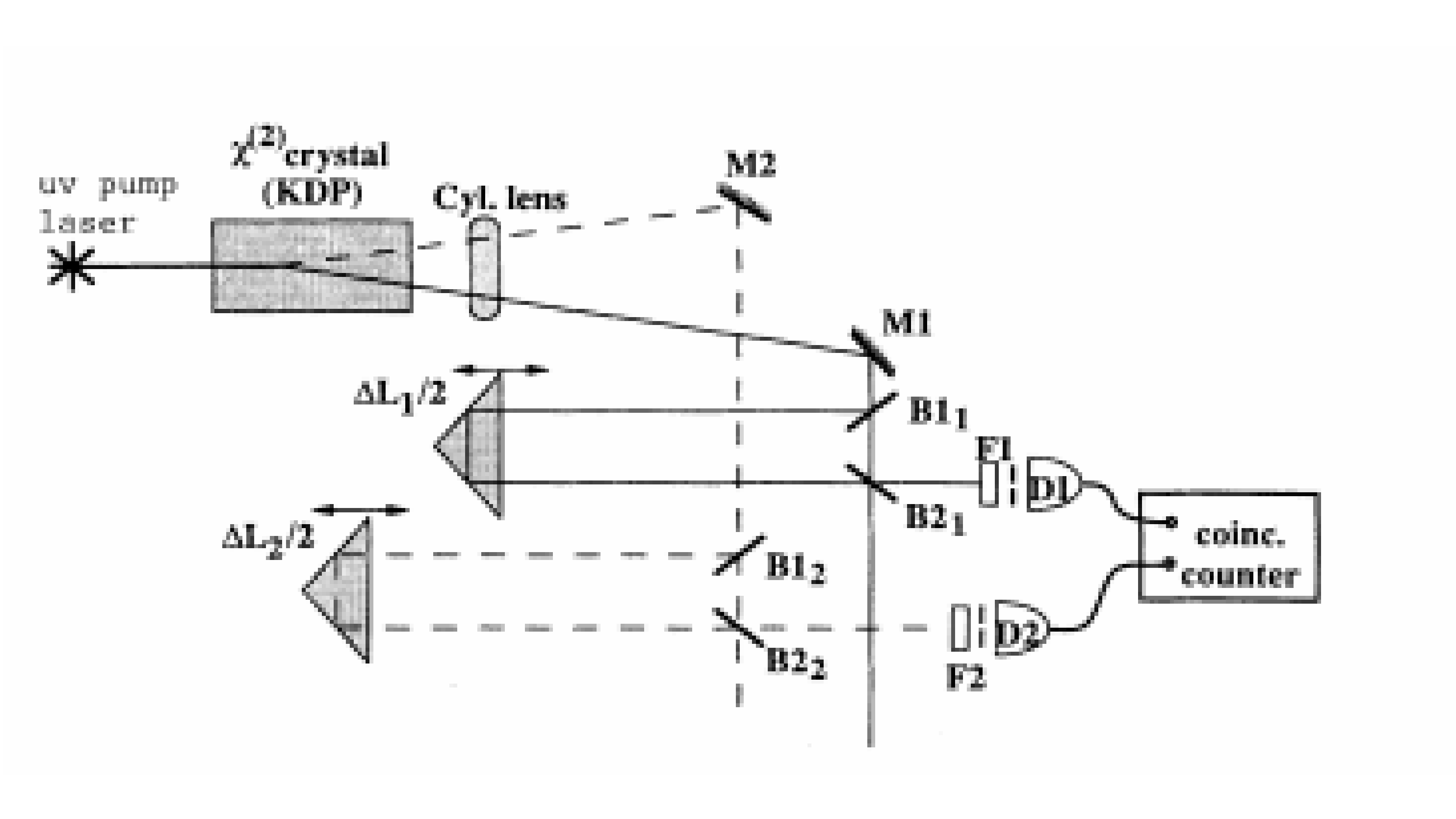}
%\end{center}
\caption{Description of the set-up of the Franson experiment described in \cite{Kwiat1}, from which the figure reproduced here has been taken. {\color{black} In this experiment only events recorded at one of the two detectors at each end of the optical device are registered. Since by symmetry considerations $p(D_A \bigcap D_B)=p(D'_A \bigcap D'_B)$, these two detectors are sufficient to test the expected fringes of interference.}}
\label{fig:FransonExp}
\end{figure}

A general review of the actual setup of the Franson experiment and other related optical tests can be found in reference \cite{Belinskii}.  Fig. \ref{fig:FransonExp} shows the setup of the Franson experiment described in reference \cite{Kwiat1}. In this experiment pairs of photons locked in time and energy are produced via parametric down-conversion by splitting photons from a single-mode laser pulse with wavelength $\lambda_p=351.1~\mbox{nm}$ inciding in a crystal possessing a $\chi^{(2)}$ non-linearity. The laser pulse has a typical time width $T \sim 20~\mbox{ns}$. 

The produced photons, $A$ and $B$, have a typical coherence time $\tau \sim 36 \ \mbox{$\mu$m}/c \sim 10^{-4}~\mbox{ns}$ and a precisely defined total energy, equal to that of the splitted incident photon:   
\begin{equation}
\label{split1}
\omega_A + \omega_B = \omega_p \equiv \frac{c}{\lambda_p},  
\end{equation}
with
\begin{equation}
\label{split2}
\omega_A \sim \omega_B \sim \frac{\omega_p}{2}, \ \ \mbox{and} \ \ \ \Delta \omega_A \sim \Delta \omega_B \sim \tau^{-1} \gg \Delta \omega_p \sim 0.
\end{equation}
The two photons are then sent in opposite directions into two unbalanced Mach-Zender-type interferometers with a longer arm and a shorter arm. The length differences between the two arms of each interferometer are set to $\Delta L_{A,B} \sim 63 \ \mbox{cm}$, which is much longer than the typical coherence length of the propagating photons,
\begin{equation}
\Delta t\equiv \frac{\Delta L_{A,B}}{c} \sim 2~\mbox{ns}  \ \ \gg \ \  \tau \sim 10^{-4}~\mbox{ns}.
\end{equation} 
Hence, there is no single-particle interference in the unbalanced interferometers. 
Yet, the time delay that these length differences introduce is much shorter than the time width of the incident laser pulse
\begin{equation}
\label{length_difference}
\Delta t\equiv \frac{\Delta L_{A,B}}{c} \sim 2~\mbox{ns} \ \ \ll \ \ T~\sim 20 \mbox{ns}.
\end{equation} 
These length differences $\Delta L_{A,B}$ can be precisely controlled and modified at each one of the unbalanced interferometers  for different settings of the experiment, but they are always set at equal values:
\begin{equation}
\label{constraint}
\Delta L_A = \Delta L_B \equiv \Delta L, 
\end{equation}
{\color{black} with high precision, 
\begin{equation}
\label{constraint2}
\left|\Delta L_A - \Delta L_B\right| \ll \lambda_p.
\end{equation}
}This length difference $\Delta L$, which is the only parameter in the described experimental setting, introduces a relative total phase 
\begin{equation}
\label{fringes_scale}
\phi_A + \phi_B = 2\pi \ \frac{\omega_p \ \Delta L}{c}, 
\end{equation}
between the splitters located at the exits of the two unbalanced interferometers. {\color{black} It is important to notice that even though the phases acquired by each one of the photons of a pair $\phi_{A, B}=2\pi \omega_{A, B} \Delta L/c$ fluctuate largely over a long sequence of repetitions of the experiment due to their limited coherence, $2\pi \ \tau^{-1} \Delta L/c \gg 1$,  under constraint (\ref{constraint2}) the total phase (\ref{fringes_scale})  remains constant to a much more stringent amount $2\pi \ \Delta \omega_p \ \Delta L/c \sim 0$.}\\

Each one of the two photons leaves its interferometer through one of the two available ports at the corresponding splitter and it is recorded by a detector, which measures its time of arrival. Pairs of photons that arrive at their respective detectors at different times (that is, non-simultaneous pairs) are defined either as event $\#3$ or event $\#4$, depending on which one of the two photons arrives earlier. These events occur with probabilities $p_3$ and $p_4$, as defined in (\ref{probabilities}), and they account for half of all pairs. 

\begin{figure}
%\begin{center}
\includegraphics[width=12cm]{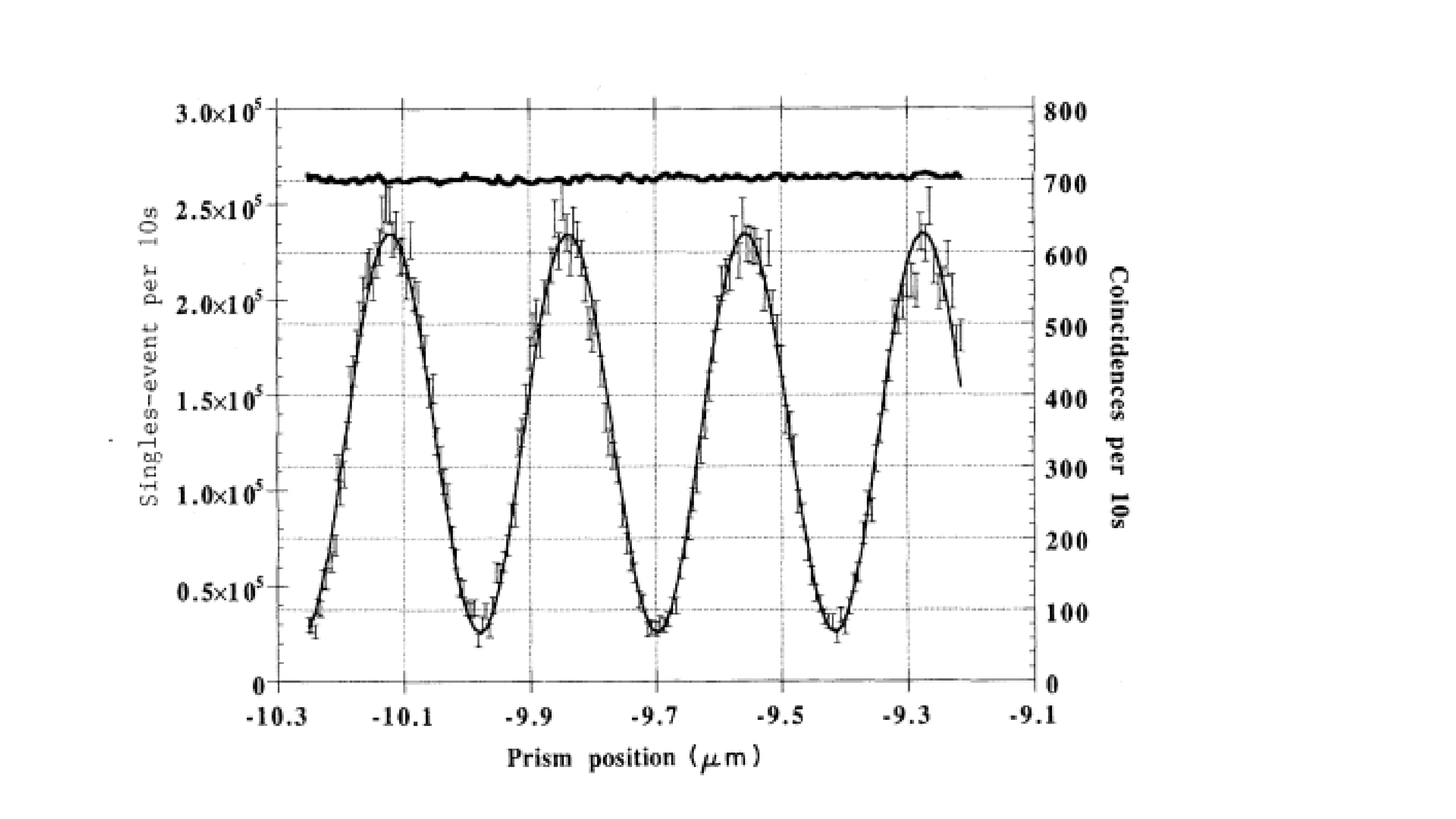}
%\end{center}
\caption{Experimental data collected in the Franson experiment described in \cite{Kwiat1}, from which this figure has been taken. The relative number of simultaneous events collected in a given pair of detectors shows a characteristic pattern of interference fringes with characteristic period of ${\it l}_p \sim \omega_p/c \sim 0.35 \mu m$, while the total number of events in each detector shows no such pattern.}
\label{fig:FransonExpData}
\end{figure}

The other half corresponds to pairs for which both photons are detected simultaneously. If they are detected either by detectors $D_A$ and $D_B$ or by detectors $D'_A$ and $D'_B$ they are counted as event $\#1$. On the other hand, if they are detected either by detectors $D_A$ and $D'_B$ or by detectors $D'_A$ and $D_B$ they are counted as event $\#2$. As shown in Fig. (\ref{fig:FransonExpData}), which has been taken and reproduced here from reference \cite{Kwiat1}, these events occur with probabilities that show a characteristic pattern of interference fringes as a function of the length difference $\Delta L$ introduced in the experimental setting (\ref{constraint}), even though the total number of photons collected at each one of the four detectors do not show any such fringes, in good agreement with the predictions (\ref{probabilities}) of quantum mechanics. 

As it can be seen from Fig. (\ref{fig:FransonExpData}) and eq. (\ref{fringes_scale}), the period of the interference pattern in these probabilities is fixed by the wavelength of the incident photon splitted via parametric downconversion at the non-linear crystal, 
\begin{equation}
\label{fringes}
{\it l} \sim \lambda_p \sim 0.35 \ \mu m.
\end{equation}
%This length scale is much smaller than all the other length scales involved in the experiment, see the Table below.

The described interference pattern in the probabilities of simultaneous events is attributed within the framework of quantum mechanics to the experimental impossibility to distinguish if the pair of simultaneous photons arrived at their detectors either both through the longer arms or both through the shorter arms of their respective interferometers. The two possibilities are undistinguishable because, as we noticed above in eq. (\ref{length_difference}), the uncertainty in their emission time caused by the time width of the laser pulse, $T \sim 20~\mbox{ns}$, is much longer than the time delay introduced by the length difference between the longer and shorter arms, $\Delta t \sim 2~\mbox{ns}$  \cite{Franson,Franson1,Kwiat1}. 
\\

\vspace{0.25in}
\hspace{-0.15in}
\captionof{table}{Characteristic time and length scales in the Franson experiment reported in \cite{Kwiat1}.}
\begin{tabular}{|l||l|l|}
\hline
 &\multicolumn{1}{l|}{ \hspace{0.0in} ~ Time scale \hspace{0.1in} }&\multicolumn{1}{l|}{ \hspace{0.0in} ~ Length scale \hspace{0.1in} }\\
\hline\hline
Laser pulse & \hspace{0.12in} $\sim$ 20 ns & \hspace{0.12in} $\sim$ 6 m\\
\hline
Arms imbalance & \hspace{0.12in} $\sim$ 2 ns & \hspace{0.12in} $\sim$ 60 cm\\
\hline
Photons coherence & \hspace{0.12in} $\sim$ $10^{-4}$ ns & \hspace{0.12in} $\sim$ 36 $\mu$m\\
\hline
Interference fringes & \hspace{0.12in} $\sim$ $10^{-6}$ ns & \hspace{0.12in} $\sim$ 0.3 $\mu$m\\
\hline
\end{tabular}
\label{table:FransonExpScales}
\\

\section{The statistical model}

In this section we describe an explicit model of local hidden variables that reproduces the predictions of quantum mechanics for the ideal Franson experiment, as summarized in eq. (\ref{probabilities}). The model closely resembles the model introduced in \cite{david,david0,david1} to reproduce the predictions of quantum mechanics for the Bell experiment. {\color{black} It also bears some similarities as well as many crucial differences, which we highlight below, with the model of hidden variables introduced by Aerts {\it et al.} in \cite{Aerts}. The comparison between the two models will help us to make clear the novel features of our model.} 

The crux of our model is the spontaneous breaking of the time-translation gauge symmetry by the hidden configurations of the pair of photons produced in the non-linear crystal. The breaking of the time-translation symmetry in this model is tantamount, as we show below, to the impossibility to set the time of emission of the photons with precision better than roughly $10\%$ of the period of the observed interference fringes (\ref{fringes}), that is, $\sim 0.03 \ \mbox{$\mu$m}/c \sim 10^{-16}~\mbox{s}$. {\color{black} At the origin of this uncertainty is a geometric phase associated with a holonomy, as intuitively illustrated in Fig.3 in \cite{david}. {\color{black} In the example shown in that figure, three parties located on the surface of a sphere cannot agree on the orientation of a tangent vector shared between them to a precision better than the geometric phase that the vector acquires when transported over the closed loop that connects the parties. In the model discussed in this paper, the three parties are the source of the pairs of photons and the detectors at both ends of the optical device, who cannot agree on the phase of the photons shared between them due to a similar holonomy.}

Since the time uncertainty associated to this holonomy is shorter by three orders of magnitude than the coherence time of the two propagating photons, see Table \ref{table:FransonExpScales}, it cannot be discarded as a a key ingredient for a succesful description within the framework of a model of local hidden variables of the pattern of inteference fringes observed in the Franson experiment. Nonetheless, Aerts {\it et al.} claim in \cite{Aerts} that "The emission time should be one of the (well-defined) variables, because if the beam splitters of, say, the right interferometer were removed, the photons would be detected solely by the detector $+1$, and the detection time $t_E$ would indicate the moment of emission", thus discarding from their considerations the possibility of spontaneous breaking of the time translation symmetry and the appearance of a holonomy.}

%\begin{figure}
%%\begin{center}
%\includegraphics[width=9cm]{Holonomy.pdf}
%%\end{center}
%\caption{Due to the holonomy of the sphere three parties located on its surface cannot agree on the orientation of the tangent %vector shared between them to a precision better than the geometric phase $\alpha$ that the vector acquires over the closed loop. 
%%The geometric phase $\alpha$ is fixed by the curvature of the sphare and the enclosed area. In the model of hidden variables for the Franson experiment discussed in this paper the three parties involved - the source of the pairs of photons and the two splitters at the ends of their corresponding unbalanced Mach-Zender interferometers - cannot agree about the phase of the photons to a precision better than $\alpha \sim 10^{-16}~\mbox{sec}\cdot \omega_p$. This time uncertainty scale is much shorter than the coherence time of the photons and, therefore, cannot be ruled out as a key ingredient of any succesful model of hidden variables for the considered experiment.
%}
%\label{fig:Holonomy}
%\end{figure}

We consider a statistical model in which the space of possible hidden configurations of the pairs of photons consists of two separated sub-populations, each one of them occurring with a probability of one half. The first sub-population accounts for events in which the two photons of the pair are detected simultaneously, that is, events $\#1$ and $\#2$ in (\ref{probabilities}), whose probabilities depend on the total phase defined by the setting of the experiment,
\begin{equation}
\label{total_phase}
\Delta \equiv \phi_A+\phi_B - \frac{\pi}{2} = 2\pi \ \frac{\omega_p \ \Delta L}{c} - \frac{\pi}{2},
\end{equation}
while the second sub-population accounts for events in which the two photons of the pair are detected at distinct times, that is, events $\#3$ and $\#4$, whose probabilities do not depend on the setting of the experiment.

The statistical space consists of an infinitely large number of possible hidden configurations distributed over the unit circle, with a density of probability given by
\begin{equation}
\label{density}
g(\varphi_A) = \frac{1}{4} \left|\sin(\varphi_A)\right|,    
\end{equation}
where $\varphi_A \in [-\pi,\pi)$ is an angular coordinate over the circle, {\it i.e.} a phase, which determines if photon $A$ will be detected either at detector $D_A$ or at detector $D'_A$, according to:
\begin{eqnarray}
{\cal S}^{(A)}={\cal S}(\varphi_A) = \left\{
\begin{array}{cccc}
D_A,     &  \hspace{0.1in} \mbox{if} & \hspace{0.1in} \varphi_A \in & [-\pi, \ 0) \\
D'_A,     & \hspace{0.1in} \mbox{if} & \hspace{0.1in} \varphi_A \in & [0, \ +\pi)
\end{array}
\right.
\end{eqnarray}
Since $g(\varphi_A)=g(-\varphi_A)$ each one of the two possibilities occurs with a probability of one half. Let us notice that the density of probability (\ref{density}) is normalized to
\begin{equation}
\int_{-\pi}^{+\pi} d\varphi_A \ g(\varphi_A) = 1.
\end{equation}

{\color{black} This coordinate is somewhat similar to the angular coordinate $\theta$ used in the model of hidden variables built by Aerts {\it et al.} in \cite{Aerts}. In their model the random variable $\theta$ is distributed uniformly over its range $[-\pi,\pi)$, but they introduce a non-uniform boundary to produce the correct distribution.} 
\\ 

{\color{black} Each one of these possible hidden configurations may appear in four possible {\it shapes},  labelled as $(\eta_A=\pm 1, \eta_B=\pm1)$, each one with a probability of $1/4$. These {\it shapes} determine if the photons are detected either at the earlier or the later time slot, according to:
\begin{eqnarray}
{\it s}^{(N)}= \left\{
\begin{array}{cccc}
\mbox{EARLY},     &  \hspace{0.1in} \mbox{if} & \hspace{0.1in} \eta_N =-1 \\
\mbox{LATE}, \  \  & \hspace{0.1in} \mbox{if} & \hspace{0.1in} \eta_N =+1
\end{array}
\right., 
\end{eqnarray}
where $N=A,B$. For example, photon $A$ of a pair with {\it shape} defined by $\eta_A=+1, \eta_B=-1$  will be detected in the later time slot, while photon $B$ of the same pair will be detected at the earlier time slot. These binary variables come instead the continuous coordinate $r \in [0, 1]$ considered  in the model discussed by Aerts {\it et al.} in \cite{Aerts}.  \\

In simultaneous events the two photons of the pair acquire  equal phases $\phi_A=\phi_B=\pi \ \omega_p \ \Delta L/c$ as they go through their respective interferometers, which add up to a total phase that depends on the setting of the experiment (\ref{total_phase}),
\begin{equation}
{\widetilde \Delta}=\eta_A \cdot \Delta =\eta_A \cdot \left(\phi_A+\phi_B-\frac{\pi}{2}\right), \ \ \mbox{if} \ \ \eta_A=\eta_B,
\end{equation}
while in not-simultaneous events the two photons acquire opposite phases $\phi_A=-\phi_B$, so that the total phase between the two does not depend on the experimental setting,
\begin{equation}
{\widetilde \Delta}=-\eta_A \cdot \frac{\pi}{2}, \ \ \mbox{if} \ \ \eta_A=-\eta_B.
\end{equation}
} 

A hidden configuration characterized by a phase $\varphi_A$ at the exit of interferometer $A$ is described at the exit of interferometer $B$ by a phase $\varphi_B$ related to the former by the coordinate transformation,
\begin{equation}
\label{transformation}
\varphi_B = L(\varphi_A,{\widetilde \Delta}),
\end{equation}
where,
\begin{itemize}
\item If  ${\widetilde \Delta} \in [0, \pi)$, 
\begin{eqnarray}
\label{Oaknin_transformation}
\hspace{-0.15in}
L(\varphi; {\widetilde \Delta}) =  
\left\{
\begin{array}{c}
\hspace{0.01in} q(\varphi) \cdot \mbox{arc-cos}\left(-\cos({\widetilde \Delta}) - \cos(\varphi) - 1 \right), \\ \hspace{0.88in} \mbox{if}  \hspace{0.1in} -\pi \hspace{0.16in} \le  \varphi < {\widetilde \Delta}-\pi, \\
\hspace{0.01in} q(\varphi) \cdot \mbox{arc-cos}\left(+\cos({\widetilde \Delta}) + \cos(\varphi) - 1 \right), \\ \hspace{0.685in} \mbox{if}  \hspace{0.05in} {\widetilde \Delta}-\pi \hspace{0.08in} \le \varphi < \hspace{0.105in} 0, \\
\hspace{0.01in} q(\varphi) \cdot \mbox{arc-cos}\left(+\cos({\widetilde \Delta}) - \cos(\varphi) + 1 \right), \\ \hspace{0.69in} \mbox{if}  \hspace{0.25in} 0 \hspace{0.18in} \le \varphi < \ {\widetilde \Delta}, \\
\hspace{0.01in} q(\varphi) \cdot \mbox{arc-cos}\left(-\cos({\widetilde \Delta}) + \cos(\varphi) + 1 \right), \\ \hspace{0.72in} \mbox{if}  \hspace{0.21in} {\widetilde \Delta}  \hspace{0.19in} \le  \varphi  < +\pi, \\
\end{array}
\right.
\end{eqnarray}
\\

\item If  ${\widetilde \Delta} \in [-\pi, 0)$, 
\begin{eqnarray}
\label{Oaknin_transformation_Inv}
\hspace{-0.15in}
L(\varphi; {\widetilde \Delta}) =  
\left\{
\begin{array}{c}
q(\varphi) \cdot \mbox{arc-cos}\left(-\cos({\widetilde \Delta}) + \cos(\varphi) + 1 \right), \\ \hspace{0.650in} \mbox{if}   \hspace{0.11in} -\pi \hspace{0.15in} \le \varphi < {\widetilde \Delta}, \\
q(\varphi) \cdot \mbox{arc-cos}\left(+\cos({\widetilde \Delta}) - \cos(\varphi) + 1 \right), \\ \hspace{0.66in} \mbox{if}   \hspace{0.30in} {\widetilde \Delta} \hspace{0.1in} \le \varphi < \hspace{0.05in} 0, \\
q(\varphi) \cdot \mbox{arc-cos}\left(+\cos({\widetilde \Delta}) + \cos(\varphi) - 1 \right), \\ \hspace{0.92in} \mbox{if}  \hspace{0.23in} 0 \hspace{0.22in} \le \varphi < {\widetilde \Delta} +\pi, \\
q(\varphi) \cdot \mbox{arc-cos}\left(-\cos({\widetilde \Delta}) - \cos(\varphi) - 1 \right), \\ \hspace{0.75in} \mbox{if}  \hspace{0.16in} {\widetilde \Delta} +\pi \hspace{0.00in} \le \varphi < +\pi, \\
\end{array}
\right.
\end{eqnarray}
\end{itemize}
and
\begin{eqnarray*}
q(\varphi) = \mbox{sign}((\varphi - {\widetilde \Delta}) \mbox{mod} ([-\pi, \pi))),
\end{eqnarray*}
with the function $y=\mbox{arc-cos}(x)$ defined in its main branch, such that $y \in [0, \pi]$ while $x \in [-1, +1]$. 
\\

\begin{figure}
\begin{center}
\includegraphics[width=13cm]{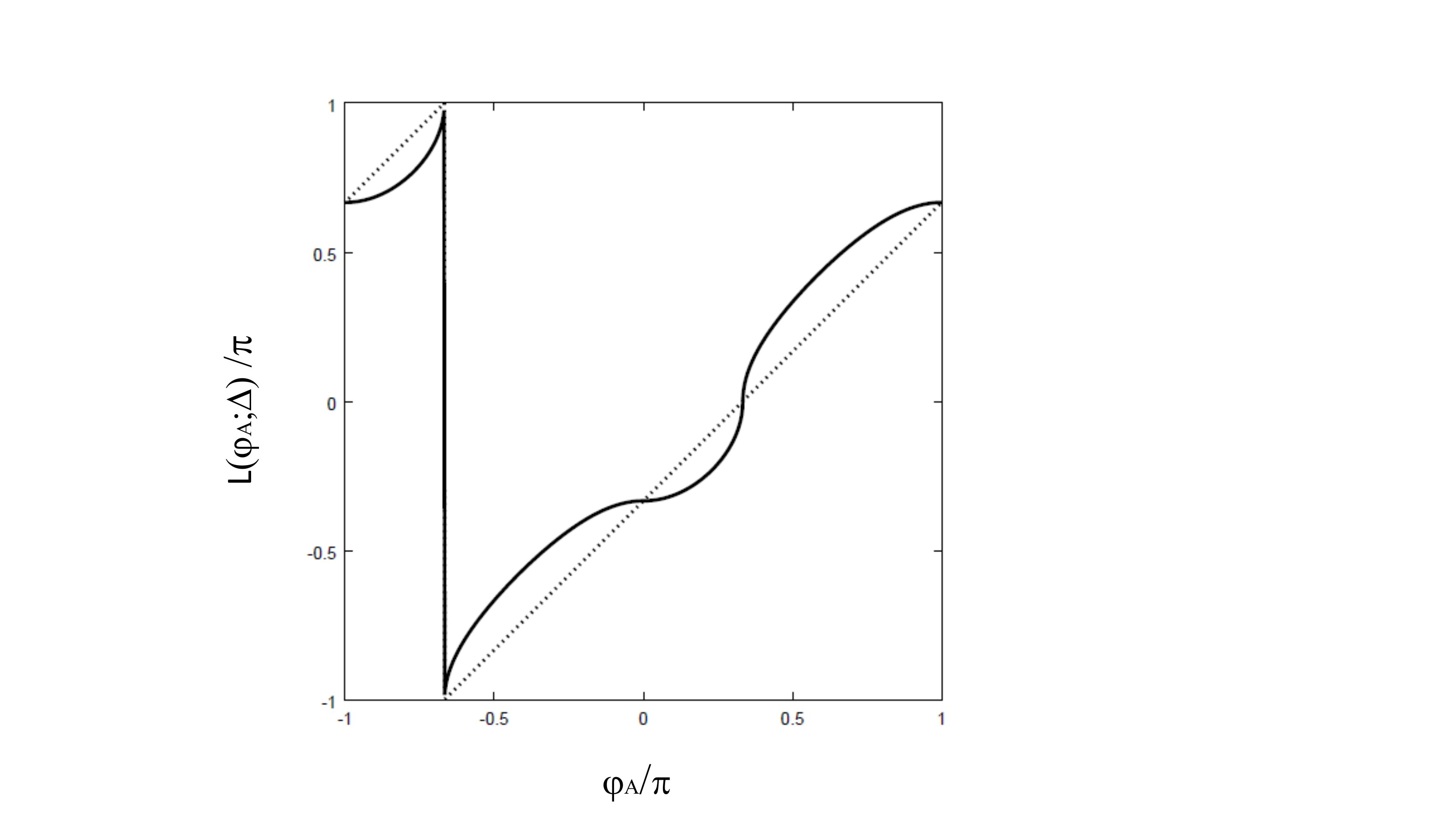}
\end{center}
\caption{Plot of the transformation law $\varphi_A \rightarrow \varphi_B = L(\varphi_A; {\widetilde \Delta}) \ \mbox{for} \ {\widetilde \Delta} = \pi/3$ (solid
line), compared to the corresponding linear transformation (dotted line).}
\label{fig:Transformation}
\end{figure}

We have shown in \cite{david,david0,david1} that this coordinates transformation fulfills the constraint
\begin{equation}
\label{free-will}
d\varphi_A \ g(\varphi_A) = d\varphi_B \ g(\varphi_B),  
\end{equation}
so that the phases $\varphi_B$ are distributed over the circle with a density of probability
\begin{equation}
g(\varphi_B) = \frac{1}{4}\left|\sin(\varphi_B)\right|    
\end{equation}
that is functionally identical to the density of probability for the phases $\varphi_A$, as it should be expected from symmetry considerations. {\color{black} Furthermore, the constraint (\ref{free-will}) states that the probability to occur of each possible configuration is independent - as it must be - from the set of coordinates used to describe them.} 

In order to keep the symmetry between the two involved parties we stipulate that photon  $B$ is detected either at detector $D_B$ or at detector $D'_B$ according to the same response function defined above for photon $A$, that is,
\begin{eqnarray}
{\cal S}^{(B)}={\cal S}(\varphi_B) = \left\{
\begin{array}{cccc}
D_B,     &  \hspace{0.1in} \mbox{if} & \hspace{0.1in} \varphi_B \in & [-\pi, \ 0) \\
D'_B,     & \hspace{0.1in} \mbox{if} & \hspace{0.1in} \varphi_B \in & [0, \ +\pi)
\end{array}
\right.
\end{eqnarray}
Therefore, for simultaneous events
\begin{eqnarray}
\nonumber
p_1 = p\left[\left(D_A \bigcap D_B\right) \bigcup \left(D'_A \bigcap D'_B\right)\right] & = & \frac{1}{4}\left[1 + \cos({\widetilde \Delta}) \right], \ \ \ \\ 
\nonumber
p_2 = p\left[\left(D'_A \bigcap D_B\right) \bigcup \left(D_A \bigcap D'_B\right)\right] & = & \frac{1}{4}\left[1 - \cos({\widetilde \Delta}) \right],
\end{eqnarray}
which exactly reproduces the probabilities (\ref{probabilities}) for events $\#1$ and $\#2$.  For non-simultaneous events, on the other hand, we get
\begin{eqnarray}
\nonumber
p_3 = p\left[\left(D_A \bigcap D_B\right) \bigcup \left(D'_A \bigcap D'_B\right)\right] & = & \frac{1}{4}, \ \ \ \\ 
\nonumber
p_4 = p\left[\left(D'_A \bigcap D_B\right) \bigcup \left(D_A \bigcap D'_B\right)\right] & = & \frac{1}{4},
\end{eqnarray}
which also corresponds to the probabilities (\ref{probabilities}) for events $\#3$ and $\#4$.

{\color{black} In order to obtain these results we notice that $L(\varphi_A; {\widetilde \Delta})$ changes sign at $\varphi_A={\widetilde \Delta}$ and at $\varphi_A={\widetilde \Delta}-\pi$ (see Fig.\ref{fig:Transformation}) and, therefore,
\begin{eqnarray}
\nonumber
p\left[\left(D_A \bigcap D_B\right) \bigcup \left(D'_A \bigcap D'_B\right)\right] 
= \int_{{\widetilde \Delta}-\pi}^{0} d\varphi_A \ g(\varphi_A) + \int_{{\widetilde \Delta}}^{\pi} d\varphi_A \ g(\varphi_A), \ \ \ \\ 
\nonumber
p\left[\left(D'_A \bigcap D_B\right) \bigcup \left(D_A \bigcap D'_B\right)\right]  
= \int_{-\pi}^{{\widetilde \Delta}-\pi} d\varphi_A \ g(\varphi_A) + \int_0^{{\widetilde \Delta}} d\varphi_A \ g(\varphi_A).
\end{eqnarray}
}

{\color{black} 
The coordinate transformation (\ref{transformation}) as defined in (\ref{Oaknin_transformation}-\ref{Oaknin_transformation_Inv}) introduces the holonomy responsible for the time uncertainty mentioned above. This non-linear transformation generalizes the linear transformation ${\widetilde \varphi}_B=\varphi_A+{\widetilde \Delta}$ assumed as an unavoidable must in the model of hidden variables discussed by Aerts {\it et al.} in \cite{Aerts}. In Fig. \ref{fig:Transformation} the transformation (\ref{transformation}) is plotted against the linear transformation for the particular value ${\widetilde \Delta}=\pi/3$ for the sake of illustration. The maximum difference between the actual transformation $L(\varphi;{\widetilde \Delta})$ and the linear transformation, which bounds the geometric phase that can be accumulated in a cycle, is roughly a $10\%$ of the period of the transformation, that is, $\sim 0.1 \omega_p^{-1} \sim 10^{-16}~s$.}
\\

As already noticed, this model closely resembles the model of local hidden variables introduced in \cite{david,david0,david1}. The crux of both models is the spontaneous breaking of a gauge symmetry by the hidden configuration of the described pairs of photons, which acquires a non-zero geometric phase through certain cyclic transformations. In the model discussed here the spontaneously broken gauge symmetry is the time-translation symmetry, or equivalently the rotational symmetry of the phases $\varphi_A$, $\varphi_B$ of the hidden configurations, which cannot be described at once with respect to the two splitters and the source of the photons due to the holonomy of the model.
\\

{\color{black} In both cases, however, the gauge symmetries are statistically restored when considered over the whole population of possible hidden configurations, in agreement with Elitzur's theorem that forbids any gauge-dependent magnitude to get a non-invariant expected value \cite{Elitzur1975}. Thus,  in the model discussed in this paper the expected probabilities (\ref{probabilities}) predicted by the model depend only on the well-defined physical parameter $\phi_A+\phi_B$ that describes the experimental setting.} 

%The transformation can be readily extended to include the sub-population of non-simultaneous events by re-defining the relative %angle as
%\begin{equation}
%\Delta = (\phi_A+\phi_B) \cdot exp(-10 |t_b-t_A|^2/(\Delta t)^2),
%\end{equation}  
%so that for simultaneous events, for which $|t_B-t_A|\sim 10^{-4}\mbox{ns}$, we recover the definition given above, while for
%non-simultaneous events, for which $|t_B-t_A|\sim  \Delta t$, we have $\Delta\simeq 0$. In order to build a more precise model it %would be necessary to further explore the collected experimental data for non-simultaneous events, which will be done %elsewwhere. The aim of this paper is only to provide a proof of principle. 

\section{Discussion}

We have presented an explicit model of local hidden variables that reproduces the predictions of quantum mechanics for the Franson experiment with pairs of photons locked in time and energy produced by parametric down-convesion in a non-linear crystal \cite{Franson,Kwiat1}. The model resembles closely the model of hidden variables  introduced in \cite{david,david0,david1} to reproduce the predictions of quantum mechanics for the Bell experiment with pairs of photons prepared in a singlet polarization state. 

The crux of both models is the spontaneous breaking of a gauge symmetry by the hidden configuration of the pairs of photons, which acquire a non-zero geometric phase through certain cyclic coordinates transformations. In the model presented in this paper the broken gauge symmetry is the time-translation symmetry. The symmetry is broken at the scale of $\sim 10^{-16} s$, which is roughly the inverse of the frequency of the photon splitted via parametric down-conversion and much shorter than the coherence time of each one of the photons of the resulting pair.

%In order to get a further understanding of the symmetry breaking in a Franson experiment it would be helpful to further explore the collected experimental data beyond the binary definition of the outcomes for each pair as 'simultaneous' or 'non-simultaneous' and instead considering the actual continuous values for the time interval ellapsed between the detections of the two photons of the pair. It could also be interesting to explore the parameter space for the experimental setting described in Fig. \ref{fig:FransonExp} beyond the constraint (\ref{constraint}), and allow $\Delta L_A$ and $\Delta L_B$ to be slightly different, {\color{blue} $|\Delta L_A - \Delta L_B| >  c \cdot t_{rsp} \sim 3~\mbox{cm}$, where $t_{rsp} \sim 0.1~\mbox{ns}$ is the typical response time of the detectors, while still in the range $\Delta L_A, \Delta L_B \sim 60~\mbox{cm}$ in order to prevent single-particle interference at the exit of the unbalanced Mach-Zender intereferometers}. \\

The insight described here for pairs of photons produced via parametric down-conversion at a non-linear crystal might also be applied to study pairs of photons (or Z-bosons) produced in the decay of scalar massive particles like, for example, positronium or neutral pions \cite{Yongram}.
\\

\end{document}